# It Takes Two to Tango, but More to Assess Systemic Risk: Credit Networks Through the Lens of Hypergraphs

**Federico D. Forte***

**Abstract**

This paper provides the first analysis of credit relationships between financial institutions and firms through the lens of hypergraphs. Unlike traditional network approaches, which rely on pairwise connections, this framework explicitly represents the shared exposure of multiple financial institutions to the same firm as a simultaneous multilateral relationship. The approach is applied empirically to Credit Registry data from the Central Bank of Argentina, covering the period from August 2023 to December 2025 and focusing on commercial loans between banks and firms. Traditional centrality metrics are compared with hypergraph-specific measures to identify systemically relevant institutions. The paper also proposes an adjusted version of H-eigenvector centrality that nonlinearly weights both the centrality of neighboring institutions and each creditor's lending amount, in order to assess the relevance of a bank within the network. The systemic impact of shocking the top-ranked institutions according to each centrality metric is then estimated through an adaptation of the DebtRank algorithm. The results show that the proposed framework identifies institutions with greater shock-amplification capacity, providing a complementary tool for financial supervision and regulation.

* Present affiliation: BBVA Research, Banco BBVA Argentina. The views expressed in this publication are those of the author and do not necessarily reflect the position of BBVA. E-mail: federico.forte@bbva.com.

## 1. Introduction

Credit relationships constitute one of the fundamental pillars of financial systems. They are the core activity of banks and usually account for the largest share of their assets. At the same time, credit is one of the main drivers of firms' investment and growth, and consequently of economic activity in general. For this reason, the relationships between banks and firms often represent one of the most relevant forms of economic interconnection for the mobilization of resources in modern economies (De Masi & Gallegati, 2012).

This paper analyzes credit relationships between financial institutions and firms in Argentina through the lens of complex networks and graph theory. More specifically, to the best of our knowledge, it provides the first application of *hypergraphs* to bank-firm credit registry data for systemic risk assessment, given that they offer a natural representation of the underlying structure of interconnections in credit networks. Traditionally, network analysis has focused on the study of bilateral relationships between pairs of agents, and graphs have proven to be a suitable tool for examining the topological structure of such interactions (Barabási, 2016; Newman, 2010). Recently, however, a growing body of literature has argued that, in many empirical settings, it is more appropriate to incorporate interrelations involving more than two agents, which may connect three, four, or more agents simultaneously (for example, Battiston et al., 2020; Bianconi, 2021; Kovalenko et al., 2022). These are referred to as higher-order interactions, and hypergraphs constitute one of the main analytical frameworks used to represent them (Figure 1). Hypergraphs generalize standard graphs by allowing for the inclusion of multilateral connections, known as *hyperedges* or *hyperlinks*. This approach has been applied to academic collaboration networks (Patania et al., 2017), biochemical networks (Traversa et al., 2023), and social networks (Iacopini et al., 2024), among other examples. However, its application to financial topics remains limited, with only a few related studies using this methodology for stock-price prediction (Fang et al., 2025; Han et al., 2023), stress testing in financial markets (Akgüller & Balcı, 2026) and the assessment of intermediaries' substitutability in the sterling money market in 1906 (Accominotti et al., 2023). Thus, this paper contributes to the literature by applying a higher-order network framework to modern bank-firm credit relationships.

When a firm receives loans from multiple banks at the same time, the traditional representation in network theory has modeled these relationships as a set of bilateral connections (De Masi & Gallegati, 2012; Fujiwara et al., 2009). However, the co-financing of a single agent by multiple financial institutions actually generates a simultaneous multilateral relationship of shared exposure among many entities, rather than simply the sum of bilateral connections between them.

**Figure 1. Illustrative example of a bipartite network and its associated hypergraph**

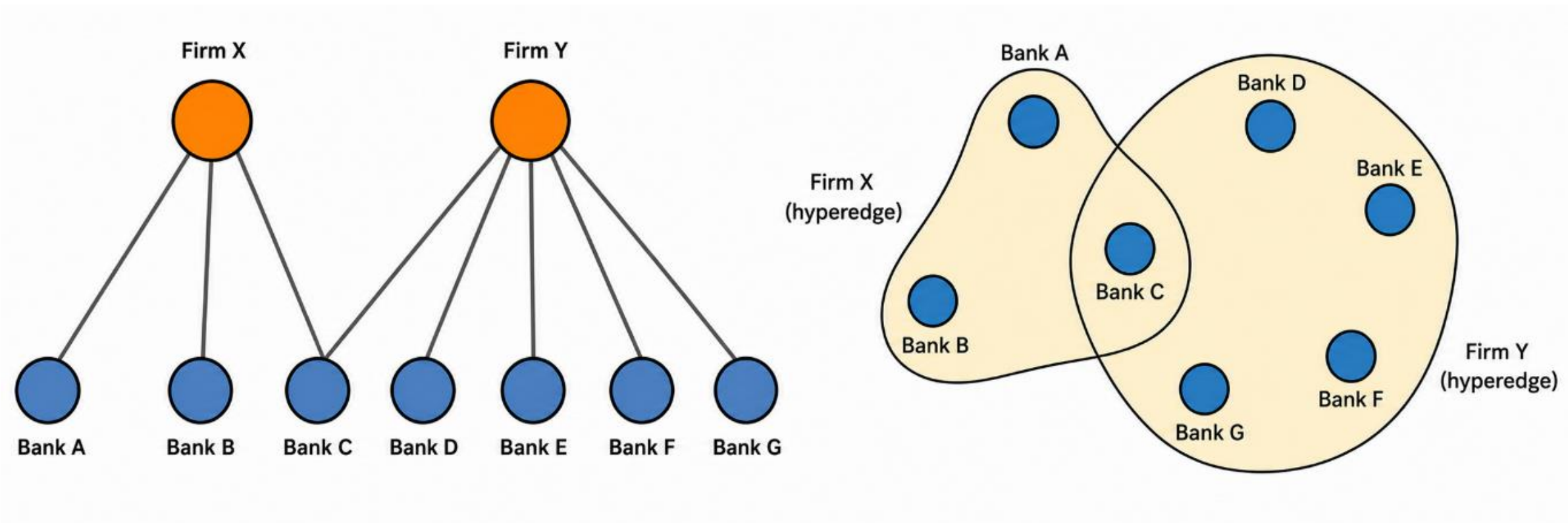

Notes: Left: illustrative example of a bipartite credit network between banks and firms. Right: hypergraph associated with the original bipartite network.

Credit networks have so far been studied within graph theory mainly through two approaches. The first represents them as bipartite networks composed of two distinct types of nodes, such as banks and firms (Marotta et al., 2015; Miranda & Tabak, 2013; Straka et al., 2018); The second relies on one-mode projections, where the original bipartite structure is projected onto one type of agents to analyze the implicit relationships among them, mediated by their shared connections to nodes in the other set. For example, banks may be linked indirectly when they are commonly exposed to the same borrowing firms (De Masi et al., 2011). Nevertheless, both approaches are based on *dyadic* representations of credit relationships and therefore simplify the *multilateral* interdependencies that arise in practice.

By contrast, the approach proposed in this paper provides a more realistic framework for capturing the potential multilateral effects of stress events in bank credit networks. Representing these inherently collective structures as a collection of bilateral ties may obscure nonlinear amplification mechanisms that emerge from the joint involvement of multiple institutions in the same credit relationships. We therefore argue that examining credit networks through the lens of hypergraphs can reveal patterns and systemic effects that cannot be fully captured by more simplified representations.

In the specific case of credit networks, when multiple banks lend to the same firm, they are jointly exposed to the same counterparty risk. A default or repayment difficulties by a firm would thus generate simultaneous stress for all its lending financial institutions. Conversely, systemic stress may also propagate through firms' roll-over risk. If a bank faces liquidity constraints and curtails credit to its borrowers, these firms may encounter difficulties in meeting their obligations to other lending banks, thereby amplifying the initial shock and potentially giving rise to iterative spillovers across the financial system. This type of systemic risk is particularly relevant in commercial credit markets, where exposures are typically larger and more concentrated than in retail segments. For this reason, this paper focuses on commercial credit rather than on individual or retail consumer loans.

Among the main recommendations issued by the Basel Committee on banking regulation is the identification of systemically important institutions, whose failure or distress could generate risks for the rest of the financial system. Regulators around the world have incorporated concepts from complex network analysis to detect such central entities, commonly described as "too-big-to-fail" or "too-interconnected-to-fail" institutions (Haldane, 2009; Hüser, 2015). Once identified, these institutions are usually subject to differentiated regulatory/supervisory treatment in order to reduce the probability of systemic financial crises.

Multiple centrality measures have been proposed to identify systemically relevant agents in different types of networks (for an extensive review, see Chun et al., 2025) . However, this field remains comparatively less developed in the context of hypergraphs. Accordingly, this paper applies centrality measures that explicitly incorporate multilateral interrelations, represented as hyperedges, and then compares their results with those obtained from more traditional graph-based metrics. In addition, we propose a complementary adjustment to an existing hypergraph centrality measure in order to better capture specific features of financial credit networks. This new adjusted measure seeks to capture the relevance of financial institutions by combining two dimensions: their total lending volume within the network and the centrality of their neighbors, while accounting for the nonlinearities inherent in multilateral interactions.

To evaluate the implications of alternative centrality measures, we estimate the systemic impact associated with the distress of the most central financial institutions identified by each of the eleven centrality metrics considered in this paper. This impact is computed through an adaptation of the *DebtRank* algorithm developed by Battiston et al. (2012), which we extend to the context of credit hypergraphs.

Our main result is that the adjusted hypergraph centrality measure proposed in this paper identifies the group of entities that generate the largest amplified systemic impact, measured as the share of total credit affected in the system. This finding has relevant implications for financial supervision and regulation, as it offers a complementary framework for analyzing interactions among financial agents and modeling shock amplification. In particular, it makes it possible to identify central entities that may not be detected by more traditional approaches.

The remainder of the paper is organized as follows. The next section describes the credit network under analysis and the database used to construct it. The third section presents the main structural properties of the network from a hypergraph perspective. Section 4 introduces the hypergraph centrality measures considered in the paper, compares them with more traditional indicators, proposes an adjustment tailored to the context of financial credit hypergraphs, and reports the resulting institution rankings. In Section 5 we estimate the systemic impact of stressing the most central agents identified by each measure. Finally, some concluding remarks are outlined in Section 6. The main results show that the financial institutions detected through the hypergraph approach generate a larger shock-amplification effect than those identified by traditional centrality measures, suggesting that this framework may provide regulators and supervisors with complementary tools to identify systemically relevant agents in credit networks.

## 2. The commercial credit network in Argentina

The main data source used in this paper is the Credit Registry of the Central Bank of Argentina (BCRA), covering the period from August 2023 to December 2025. We focus on loans classified as "commercial", defined as loans granted to firms in connection with their productive activities. Accordingly, personal, mortgage, and consumer loans are excluded from the analysis.

The BCRA suggests a reference threshold for banks to classify loans as "commercial", which is equivalent to twice the maximum annual sales value defined for the "Microenterprise" category in the Commerce sector, as established by the Secretariat for Small and Medium Enterprises (which, at the time this paper was written, operated under the Ministry of Economy). As of December 2025, this maximum sales value was ARS 1,371,080,000[1] (implying a commercial-loan reference threshold of ARS 2,742,160,000), approximately USD 1 million at the exchange rate prevailing at that moment. Although this threshold is not mandatory, and banks may classify smaller exposures within the commercial portfolio, it provides a general reference for the financial system.

For this reason, loans are included in the examined network only when the total credit granted to a given company exceeds a minimum threshold equivalent to the December 2025 reference amount. This threshold is adjusted backward for inflation, keeping it constant in real terms throughout the sample period. After applying this criterion and excluding firms with lower outstanding credit, the resulting network captures, on average, 80.3% of total credit to firms. Therefore, it covers a substantial and representative share of the loan universe under examination.

On average, 4132 ± 917 firms[2] and 157 ± 14 financial institutions were active each month during the period analyzed (Figure *2*, left), connected through 28,326 ± 5,567 credit relationships per month. The sample includes more than six types of financial institutions (Table *1*), but banks account for 98.5% of the credit granted in this commercial segment. Therefore, in the following sections, unless another type of entity is explicitly mentioned, we use the terms "financial institutions" and "banks" interchangeably.

[1] Resolution 54/2025 of the "Secretariat for Small and Medium Enterprises, Entrepreneurs and the Knowledge Economy" under the Ministry of Economy.

[2] It should be noted that these figures refer only to firms whose total debt exceeds the reference threshold described above, whereas the complete universe of companies with debt in the financial system exceeds 200,000.

Figure 2. **Monthly credit networks: number of nodes and average degree**

Note: Left: number of lending financial institutions and firms with debt in the financial system (above the minimum threshold). Right: average degree of firms, i.e., the average number of banks from which each firm has borrowed, and of financial institutions, i.e., the average number of firms to which each institution provides credit.

The most basic way to analyze this database using graph theory is through the concept of a bipartite network. This structure consists of two disjoint sets of nodes (financial institutions and firms), where connections can only be established between nodes of different types. Hence, the degree of a node is the number of nodes from the opposite set to which it is connected. In this context, therefore, the average degree of a bank corresponds to the average number of firms to which it lends, while the average degree of a firm corresponds to the number of financial institutions from which it receives credit. In the Argentine case, each debtor firm borrowed, on average, from 7 institutions over the period under study, with low variance (Figure *2*, right), while each financial institution provided credit to an average of 179 firms. This latter figure exhibits substantial dispersion: banks had an average degree of 368, whereas other types of institutions had significantly fewer connections (Table *1*).

Table 1. **Types of financial institutions in the Argentine commercial credit network**

| Type of financial institution | Number of institutions | Share of total credit | Average degree (firms per institution) |
|---|---|---|---|
| Banks | 69 | 98.4% | 368 |
| NFCCI | 7 | 0.7% | 323 |
| ONFCP | 60 | 0.7% | 9 |
| Others | 20 | 0.1% | 5 |

Notes: NFCCI: Non-Financial Credit-Card Issuers; ONFCP: Other Non-Financial Credit Providers (mutual associations, credit cooperatives, social clubs, etc.). Others: include mutual guarantee institutions, financial trusts, and providers of P2P lending services through platforms.

## 3. Basic topological metrics: bipartite networks, one-mode projections and hypergraphs

As mentioned above, the literature has usually addressed this type of interconnections through the concept of bipartite networks. For example, this perspective has been applied to the analysis of credit networks in Japan (Fujiwara et al., 2009), Italy (De Masi & Gallegati, 2012) and Brazil (Miranda & Tabak, 2013). Mathematically, these networks can be represented by a graph $G(V, U, E)$, composed of two disjoint sets of nodes: $V$ (banks) and $U$ (firms), with $V \cap U = \emptyset$, such that for every edge $(i, j) \in E$, it must be the case that $i \in V$ y $j \in U$, or vice versa. The matrix representation of these graphs is given by an adjacency matrix $A$, where, in its basic binary form, each element $a_{ij} = 1$ if there is a loan granted by bank $i$ to firm $j$, and $a_{ij} = 0$ if there is no credit relationship between the two nodes. It is also possible to work with a weighted matrix W, whose components $w_{ij}$ may represent, for example, the total amount of the loan granted, rather than simply indicating the existence or absence of a relationship in binary form (Figure 3 provides an example of a visualization of this type of network).

Another way to study this object is through a one-mode projection of the adjacency matrix. That is, the original graph can be projected onto the subspace composed of only one type of node (represented by the matrix $P = AA^T$), where two banks are connected if they lend to the same firm. This matrix can be binary, or it can be weighted according to, for example, the number of jointly co-financed firms or the total amount of credit involved in the common exposure. This approach is also widely used in the literature (De Masi et al., 2011; Miranda & Tabak, 2013). The projection onto the set of banks of the bipartite network shown in Figure 3 is displayed in Figure 4.

Figure 3. **Commercial credit relationships as a bipartite network**

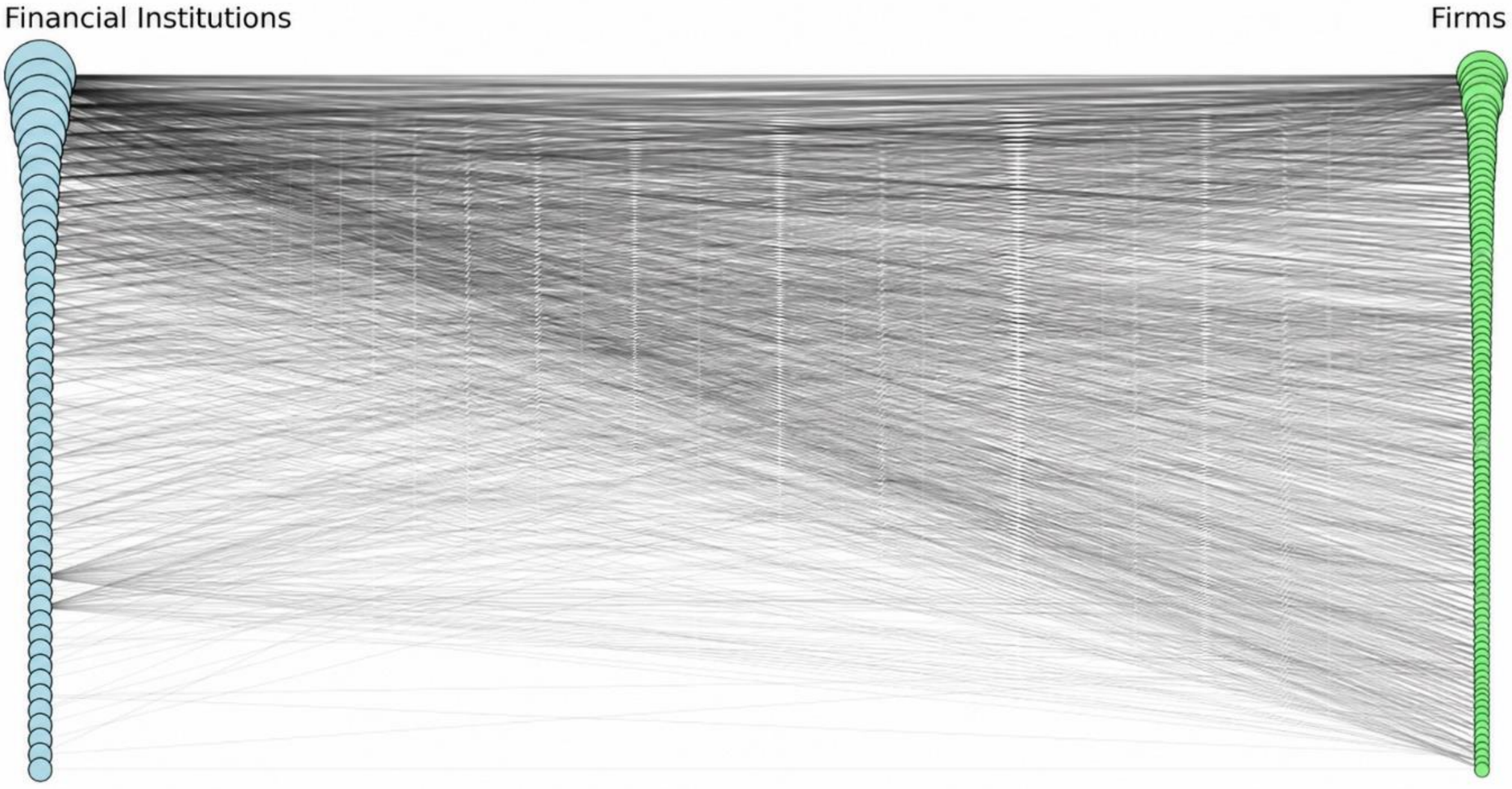

Notes: Bipartite credit network for October 2024. This month was selected because it is the one with the number of nodes closest to the average for the period under study. The size of the circles is associated with the total credit corresponding to each node. The lines represent loans granted by financial institutions (left) to firms (right), and their thickness is associated with the size of the loan. For presentation purposes, this visualization uses a subsample comprising the 50 financial institutions with the largest credit volume and the 100 largest borrowers in their joint portfolio.

Figure 4. **Bank-Bank projection of the commercial credit network**

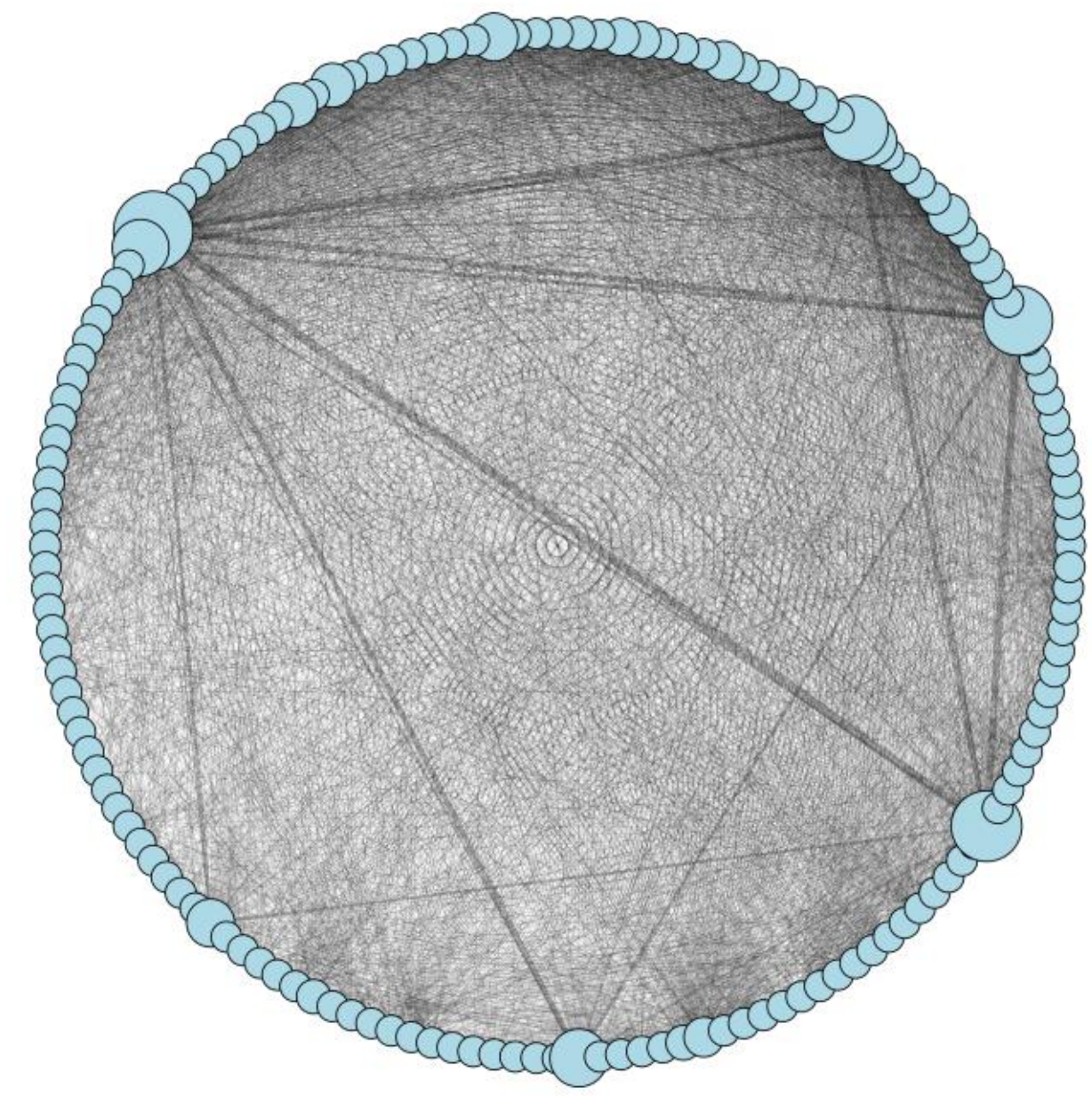

Note: Commercial credit network for October 2024. This month was selected because it is the one with the number of nodes closest to the average for the period under analysis. Node size is associated with the total amount of credit granted by each financial institution. The lines represent common credit exposures among financial institutions to the same firms, and their thickness is correlated with the total amount of joint exposure.

Based on these approaches, it is possible to compute basic topological metrics to examine the structure of interrelations in this network (Figure 5). In the case of the bipartite network, the average density over the period analyzed was 4.5% ± 0.5 p.p., a relatively low value compared with other types of networks but in line with the values observed in empirical financial networks around the world, which usually tend to be sparse (Elosegui et al., 2022; Forte, 2020; Hüser, 2015).

Figure 5. **Basic topological metrics of the bipartite network and its one-mode projection onto banks (monthly networks)**

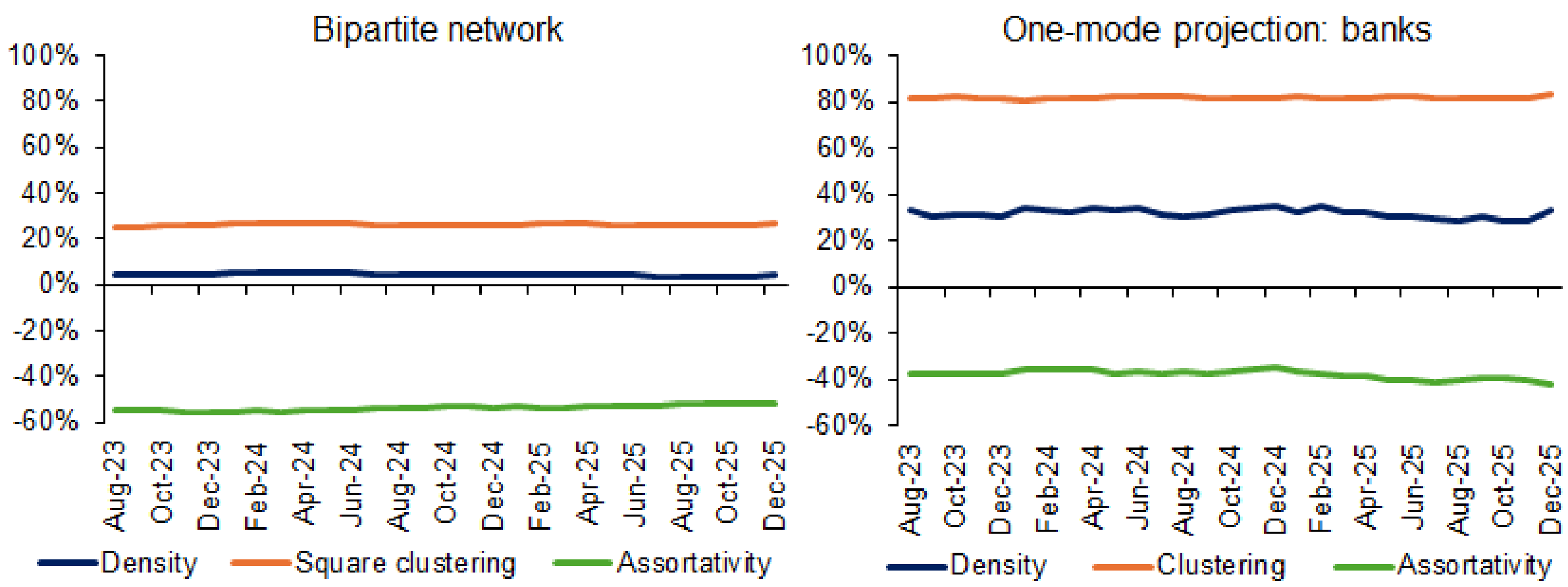

Notes: Left: metrics computed from the unweighted bank-firm bipartite networks. Right: topological metrics of the one-mode projection of the original adjacency matrix onto the set of banks.

The clustering coefficient of these networks[3] averaged 26.4% ± 0.5p.p., also in line with the values typically observed in financial networks. Finally, the network is clearly disassortative, meaning that

[3] The traditional clustering coefficient captures the share of potential triangles that are actually observed. Because triangles cannot occur in bipartite networks, we instead use the *square clustering* coefficient, which measures the prevalence of

highly connected banks tend to lend comparatively more to less connected firms than less connected banks, and vice versa. Disassortativity is a predominant feature of financial networks. None of these metrics exhibited large changes or significant volatility during the period under analysis.

When the same network is instead analyzed through its one-mode projection onto banks, attention shifts to the pattern of connections generated by banks' common exposures to the same firms (Figure 5, right). A distinctive feature of this approach is that it mechanically inflates connectivity metrics, since each firm borrowing from multiple banks generates connections among all of them. This is reflected in the average network density, which rises to 32% ± 1.9p.p, and also in the clustering coefficient (now measured as the ratio of total triangles to possible triangles), which reaches the very high value of 82% ± 0.5p.p. These connectivity levels in financial networks are usually found only in this type of projections, since in other empirical networks (e.g., interbank loans, credit, and payment networks), these metrics tend to be significantly lower. It is also worth noting that the disassortative mixing persists in this context: banks with fewer connections tend to be linked through common exposures (loans to the same firms) with highly connected banks rather than to banks with a similar degree[4].

In this paper, we propose using hypergraphs to provide a more insightful representation of the structure of interconnections in bank-firm credit networks. Hence, this type of networks can be represented as a hypergraph $H(V, E)$, consisting of a set of nodes $V = \{i_1, \dots, i_n\}$, which correspond to lending financial institutions, and a set of links $E$. The key difference with the traditional approach is that a link is now no longer restricted to a pair $(i, j)$, and, instead, each element of $E$ is itself a set that may contain two, three, or more institutions. Thus, relationships are not necessarily bilateral (although bilateral relationships remain possible) but may simultaneously involve multiple institutions. These links are known as *hyperedges*, and their *size* is determined by the number of nodes they contain. Hyperedges may also be weighted through a function $w: E \to \mathbb{R}$, where $w_i(e)$ denotes the weight of node $i$'s participation in hyperedge $e$. A hypergraph is said to be $m$-uniform when all its hyperedges have the same size $m$, although this is rarely the case in real-world applications.

Applying this approach to the credit network, when one or several firms have outstanding debt with a set of $m$ banks, this joint exposure can be represented as a hyperedge. The resulting hypergraph therefore contains only one type of node (financial institutions), while each hyperedge represents a unique set of creditor banks lending to the same firm or firms. Firms borrowing from exactly the same set of banks are thus associated with the same specific hyperedge.

In the Argentine credit network examined here, the average number of hyperedges per month was 3455±638, with a mean size of 6.9 ± 0.3 banks per hyperedge (weighted by the number of debtor firms). The modal hyperedge size was 5.3±0.5 (Figure 6, left). Thus, the most common sets of banks sharing the same exposure typically comprise between five and seven institutions, showing a substantial departure from the dyadic representation provided by conventional graphs.

The hypergraphs under analysis are far from uniform, as their hyperedge sizes range from 2 to 36 banks (Figure 6, right). As discussed in the next section, this lack of uniformity poses analytical challenges for the application of some conventional methods.

four-node cycles (that is, configurations in which two banks are both connected to the same two firms), as in Lind et al. (2005) and Luu & Lux (2018).

[4] Disassortative mixing persists even when links are weighted by loan amounts.

Figure 6. **Characteristics of hyperedges in the monthly hypergraphs**

Note: Left: Number of distinct hyperedges in each monthly hypergraph, together with their mean and modal sizes (weighted by the number of debtor firms in each hyperedge). Right: Histogram of the average monthly frequency distribution of hyperedge sizes over the period under analysis.

## 4. Identification of systemically important financial institutions

### *Bipartite networks*

One of the main contributions of network analysis and graph theory to the literature on financial systems has been the development of tools for identifying the most central institutions in a network, thereby helping to quantify and mitigate the systemic risk associated with them (Alaeddini et al., 2023; Martinez-Jaramillo et al., 2019).

The traditional centrality measures most commonly used in financial networks include node *degree* (which, in this context, corresponds to the number of firms to which a bank lends), node *strength* (measured here as the total amount lent by each bank in a given month) and *eigenvector centrality*:

$$Degree_i = \sum_{\forall j \neq i} a_{ij} \quad \text{(I)}$$

$$Strength_i = \sum_{\forall j \neq i} w_{ij} \quad \text{(II)}$$

$$Eigenvector_cent_i = x_i, \text{such that } \lambda \boldsymbol{x} = A\boldsymbol{x} \quad \text{(III)}$$

The first two measures are computed directly from the original bipartite network, as is common practice in financial supervision worldwide. By contrast, the standard definition of eigenvector centrality requires a square adjacency matrix, since the centrality value of a node $i$ corresponds to the $i$-th entry of the eigenvector associated with its leading eigenvalue. Hence, this last centrality metric can only be extracted from the one-mode projection of the original bipartite matrix, which yields a square bank-bank matrix. Specifically, the bank-bank projection is constructed from the bipartite network weighted by loan amounts in real terms.

### *Centrality measures in hypergraphs*

Under a hypergraph framework, the analysis no longer relies on adjacency matrices. Instead, the object of study becomes a higher-order tensor, or hypermatrix, with a more complex multidimensional structure. Building on recent advances in the literature that extended the notion of eigenvalues to higher-order tensors and hypermatrices (Qi & Luo, 2017), several measures have been proposed to generalize eigenvector centrality to the context of hypergraphs. One of the most prominent is known as H-eigenvector centrality (Benson, 2019). This measure generalizes the notion of eigenvectors to higher-dimensional settings. In other words, it preserves the same intuition as the original concept, whereby the centrality of a node depends on the centrality of its neighbors, while simultaneously allowing for the nonlinearities that arise in environments characterized by multilateral relationships.

Tensors provide a useful algebraic representation for hypergraphs. Broadly speaking, they generalize matrices to higher-order dimensions. Thus, given an $m$-uniform hypergraph $H = (V, E)$, where $V$ denotes the set of nodes and $E$ the set of hyperedges, an adjacency tensor (or hypermatrix) $T$, can be defined as follows:

$$T_{i_1,\dots,i_m} = \begin{cases} 1 & if\ (i_1, \dots, i_m) \in E \\ 0 & otherwise \end{cases}$$

where $T_{i_1,\dots,i_m}$ is analogous to the adjacency matrix $A$ for the bilateral case. Relationships can be defined in binary form, as in the equation above, or by incorporating a weight $w(\{i_1, \dots, i_m\})$, which in our case essentially reflects the amount lent by each of the financial entities $\{i_1, \dots, i_m\}$ within each hyperedge.

Given this tensor representation, if the hypergraph is strongly connected, there exists a centrality vector **c** that satisfies the following condition:

$$c_i^{m-1} = \frac{1}{\lambda} \sum_{i_2,\dots,i_m=1}^{n} T_{i i_2 \dots i_m} c_{i_2} \dots c_{i_m} \Rightarrow \lambda c^{[m-1]} = T c^{m-1}$$

Where $\boldsymbol{c}^{[m-1]}$ denotes the component-wise $(m-1)$-th power of **c**. The centrality vector **c** is then computed iteratively using the power method, in order to obtain the dominant H-eigenvector and, accordingly, the *H-eigenvector centrality* associated with each financial institution.

Other types of eigenvectors have been proposed in the context of hypergraphs (Benson, 2019; Qi & Luo, 2017), but we focus on the H-eigenvector for several reasons: it is less computationally demanding, it admits a unique solution under weaker assumptions than alternative approaches, and it captures the richness of the nonlinear nature of multilateral relationships in hypergraphs without reducing them to a dyadic setting (as some alternative methods do).

A difficulty that arises when applying this centrality computation to financial networks is that it requires the hypergraph to be $m$-uniform; that is, all hyperedges must contain the same number $m$ of nodes, or equivalently, all hyperedges must have homogeneous size[5]. This is not usually the case in empirical applications, and it is certainly not the case here. Moreover, if each $m$-uniform sub-hypergraph were analyzed separately, the complexity of the actual hypergraph would be lost, along with a substantial part of the interdependencies among hyperedges of different sizes within the network.

To overcome this limitation, several methodological strategies have recently been proposed. We discuss below three approaches that have been developed in the literature to address this issue:

1) The first approach proposed to quantify centrality values in non-uniform hypergraphs was the notion of *vectorial centrality score*. Instead of just a single centrality value assigned to each node, a centrality vector is constructed for each node $i$, $\boldsymbol{v}^i = \mathbb{R}^{m-1}$, where each component of the vector represents the centrality value of that node in the corresponding $m$-uniform sub-hypergraph. A setback of this solution is that, once the hypergraph is decomposed according to the size of its hyperedges, the resulting sub-hypergraphs may fail to be strongly connected. Moreover, this decomposition disregards part of the higher-order interdependence among hyperedges of different sizes that characterizes the original network structure. Hence, Kovalenko et al. (2022), in order to address that limitation, suggested using the incidence matrix of the hypergraph (whose rows correspond to nodes and the columns correspond to hyperedges), and projecting it onto the set of hyperedges. This yields a square matrix, namely the *line graph* derived from the hypergraph. On this object, the centrality of the edges can be computed using traditional procedures, such as the classical eigenvector centrality. Once the centrality of each hyperedge -$c(h)$- has been computed, the authors suggest defining node centrality as the sum of the centralities of the hyperedges in which the node participates. More specifically, they propose constructing, for each node $i$, a normalized vector $\boldsymbol{c_i} = \mathbb{R}^{m-1}$, where each of its $k$ components are given by:

$$c_{ik} = \frac{1}{k} \sum_{\substack{i \in h \in E \\ |h| = k}} c(h)$$

[5] This requirement is analogous to the traditional square-matrix condition in standard spectral analysis to compute eigenvalues, extended to the tensor setting through the need for an $m$-uniform hypergraph.

Where $c(h)$ denotes the centrality value of hyperedge $h$, and $|h|$ refers to the size of that hyperedge. The authors show that this vector is normalized.

2) A second proposal was put forward by Tudisco & Higham (2021), who introduce a family of spectral centrality measures for non-uniform hypergraphs, which allows a simultaneous computation of the relative importance of both the nodes and hyperedges. The central idea is to extend the logic of eigenvector centrality in traditional graphs to settings in which interactions are not necessarily bilateral, but rather of higher order. Under this notion, in a simple graph a node is central if it is connected to other central nodes. In a hypergraph, by contrast, a node may participate in hyperedges that group several nodes at once. The authors therefore propose a mutually reinforcing relationship: a node is important if it participates in important hyperedges, and a hyperedge is important if it contains important nodes. This formulation makes it possible to capture the higher-order interaction structure without necessarily reducing the hypergraph to a clique projection or to a network of pairwise relations.
   Starting from the incidence matrix of the hypergraph, Tudisco and Higham formulate a nonlinear eigenvalue problem that generalizes both traditional eigenvector centrality and some tensor-based measures for uniform hypergraphs. The flexibility of the approach arises from the choice of certain nonlinear functions that determine how node centralities are aggregated within each hyperedge and how hyperedge centralities are aggregated at each node. They study three main specifications: a linear version, similar to eigenvector centrality applied to the bank projection and to the line graph; a "log-exp" version, which generalizes tensor-based Z-eigenvector centrality; and a "max" version, which assigns high centrality to nodes that participate in at least one highly important hyperedge. The result is a family of three measures that can be adapted to different assumptions about how centrality is transmitted within group interactions (see Appendix A for further details on the mathematical formulation).

3) Finally, the most recent proposal (and the one discussed in greater depth later in this paper), is that of Contreras-Aso et al. (2024). They propose a "uniformization" method for making hyperedge sizes equal in hypergraphs that are originally non-uniform. Once the baseline hypergraphs have been "made" uniform, H-eigenvector centrality measures can be computed in the same way as in the benchmark case described above. The main idea they propose is to "expand" the size of smaller hyperedges by adding "artificial auxiliary nodes" weighted in such a way that their impact on the centrality metrics is neutral. They refer to this procedure as the "*uplift*" of smaller hyperedges. In addition, they also propose "shrinking" larger hyperedges to a smaller size, such as $p$. To do so, they suggest decomposing ("*projecting*") each hyperedge of size $k > p$ into all possible $C(k,p) = \binom{k}{p}$ combinations of smaller hyperedges of size $p$ (see Appendix B for further details on the mathematical formulation of this proposal).

To compute this third type of centrality using the method proposed by Contreras-Aso et al., it is necessary to define the size to which all hyperedges in the hypergraphs under analysis should be "uniformized". In this paper two alternative strategies are proposed: (1) setting all hyperedges to the modal size, by uplifting the smaller hyperedges and projecting the larger ones; and (2) increasing the size of all hyperedges to 19 through the uplift procedure (this level is chosen because hyperedges of up to this size account for at least 80% of the total amount lent in the networks for all months under analysis) and discarding hyperedges larger than that value, since expanding all hyperedges to the maximum observed size (36) would exponentially increase the number of artificial nodes required, while adding little information, as only a relatively small share of total loans is concentrated in that segment of larger hyperedges.

In this paper we propose a complementary refinement of the algorithms developed by Contreras-Aso et al. (2024) for the computation of centrality in financial hypergraphs. The original formulation calculates the centrality of each node $i$ as a function of the centrality of its neighboring nodes within the hyperedges in which it participates. In the context of credit networks, the standard weighted version of this measure would assign to each interaction the total weight of the shared hyperedge. Here, we keep that idea but weight the centrality of those neighboring nodes by the amount that the node $i$ lends in each hyperedge in which it participates.

This adjustment avoids the potential bias of identifying as relevant those nodes that, for example, participate in large transactions alongside highly central banks while contributing only small amounts, or small shares of their own portfolios, which may not adequately reflect their core activity. If neighbors' centrality is not weighted by the node's own contribution, the measure may overstate the relevance of small or secondary banks whose relevance derives mainly from participating with minor amounts in transactions largely driven by more central banks.

Mathematically, the complementary refinement we propose for computing this weighted version of the H-eigenvector centrality can be formulated as follows:

- The central idea behind the *traditional* H-eigenvector centrality is that the centrality $c_i$ of node $i$ depends on the centrality of the neighboring nodes with which it shares hyperedges. Since more than one neighbor is shared within each hyperedge (more precisely, $m-1$ neighbors in uniform hypergraphs), this formulation quantifies the joint centrality of the neighboring nodes, $c_j$, multiplicatively:

  Standard unweighted H_eigenvector centrality: $\lambda c_i^{m-1} = \sum_{i \in e \in E} \prod_{\substack{j \in e \\ j \neq i}} c_j$

- The weighted version of this centrality measure, where weights are given by the amount of credit involved in each hyperedge, can be written as:

Standard weighted H_eigenvector centrality: $\lambda c_i^{m-1} = \sum_{i \in e \in E} w_e \prod_{\substack{j \in e \\ j \neq i}} c_j$

Where $w_e$ denotes the total amount of credit involved in each hyperedge.

- However, in the context of financial networks, we argue that identifying central agents in credit networks requires more than examining the centrality of the neighbors with which a bank shares credit transactions and the total amount associated with the hyperedge. It is also relevant to account for how much each entity contributes to these joint exposures. Otherwise, small or relatively marginal entities may be classified as important merely because they participate in large hyperedges with highly connected banks, even if they consistently do so with small amounts that are not systemically relevant. For this reason, we propose weighting separately the contribution of each entity to the corresponding hyperedge, thereby disaggregating the weight assigned to each node involved in the shared exposure. Mathematically, our proposal is the following:

  H_eigenvector centrality weighted by individual contribution to each hyperedge:

$$\lambda c_i^{m-1} = \sum_{i \in e \in E} h_{ie} \prod_{\substack{j \in e \\ j \neq i}} h_{je} c_j$$

  Where $h_{ie}$ is the amount lent by node $i$ in hyperedge $e$. In summary, the proposed centrality measure multiplicatively incorporates both the amount lent by each entity in each hyperedge and the centrality of the nodes sharing those hyperedges. This formulation therefore combines the relevance of each node's neighbors, the importance of the hyperedges in which it participates, and the individual contribution of node $i$ to each corresponding hyperedge.

To summarize, Table *2* reports all the centrality metrics that will be applied to the credit network under analysis, with the aim of quantifying, through alternative methods, the relative relevance of the central agents in the system.

After computing the centrality rankings of financial institutions for all months in the sample, we compare the results using Kendall rank correlations and the top-20 overlap of entities identified by each of the metrics described in Table *2*.

Based on the average Kendall correlations (shown in Figure *7*), the first result that emerges is that, in general, all rankings obtained from the different centrality measures display positive and relatively high correlations. The lowest correlation among all possible pairs is 0.59, while the overall average pairwise rank correlation is 0.73.

Nonetheless, there are some groups of metrics that are comparatively more similar to one another. The results derived from the hypergraph-based indicators proposed by Tudisco & Higham (2021) (*HIP_TH_lin*, *HIP_TH_logexp*, and *HIP_TH_max*) entail generally the closest results to traditional metrics based on bipartite graphs (such as degree and strength), or on the bank-bank projection (*ONE_eigenvector*). In other words, these metrics tend to provide less additional information relative to the more traditional approaches.

One step below in terms of correlations with traditional metrics, is Kovalenko et al's vectorial centrality measure (*HIP_vec_cent*). By contrast, the metrics that display the lowest correlations with traditional measures are those obtained from the computation of H-eigenvectors (*HIP_HE_uplift* and *HIP_HE_upproj*). This would suggest, *a priori*, that the latter may be the measures with the greatest potential to provide new information relative to the approaches traditionally applied through graphs with dyadic connections.

Table 2. **Summary of the centrality measures computed in this paper**

| | Graph | Centrality | Abbreviation | Description |
|---|---|---|---|---|
| (I) | Bipartite | Degree | *BIP_degree* | Number of loans granted |
| (II) | Bipartite | Strength | *BIP_strength* | Total amount lent |
| (III) | One-mode projection (banks) | Eigenvector | *ONE_eigenvector* | The centrality of a node depends on the centrality of its neighbors |
| (IV) | Non-uniform hypergraph | Vectorial Centrality | *HIP_vec_cent* | Sum of the centrality of the hyperedges in which each node participates |
| (V) | Non-uniform hypergraph | Tudisco & Higham (2021) - linear model | *HIP_TH_lin* | A node is central if it participates in important hyperedges |
| (VI) | Non-uniform hypergraph | Tudisco & Higham (2021) - logexp model | *HIP_TH_logexp* | Node centrality depends on the product of the centralities of its hyperedge neighbors |
| (VII) | Non-uniform hypergraph | Tudisco & Higham (2021) - max model | *HIP_TH_max* | A node is central if it belongs to at least one highly important hyperedge |
| (VIII) | Uniformized hypergraph | H-eigenvector - uplift to maximum size | *HIP_HE_ uplift_trad* | Generalization of ONE_eigenvector to uniform hypergraphs |
| (IX) | Uniformized hypergraph | Same as (VIII), weighted by individual amount lent | *HIP_HE_ uplift_adj* | Weights neighbors' centrality and the amount lent in each hyperedge |
| (X) | Uniformized hypergraph | H-eigenvector - uplift & projection to modal size | *HIP_HE_ upproj_trad* | Generalization of ONE_eigenvector to uniform hypergraphs |
| (XI) | Uniformized hypergraph | Same as (X), weighted by individual amount lent | *HIP_HE_ upproj_adj* | Weights neighbors' centrality and the amount lent in each hyperedge |

Figure 7. **Kendall correlations between bank rankings across centrality metrics**

| | BIP_ degree | BIP_ strength | ONE_ eigenvector | HIP_ vec_cent | HIP_TH_ lin | HIP_TH_ logexp | HIP_TH_ max | HIP_HE_ uplift_trad | HIP_HE_ upproj_trad | HIP_HE_ uplift_adj | HIP_HE_ upproj_adj |
|---|---|---|---|---|---|---|---|---|---|---|---|
| BIP_degree | 1.00 | 0.74 | 0.72 | 0.79 | 0.72 | 0.74 | 0.74 | 0.70 | 0.75 | 0.68 | 0.69 |
| BIP_strength | 0.74 | 1.00 | 0.83 | 0.69 | 0.83 | 0.99 | 0.94 | 0.63 | 0.66 | 0.66 | 0.65 |
| ONE_eigenvector | 0.72 | 0.83 | 1.00 | 0.78 | 1.00 | 0.82 | 0.87 | 0.59 | 0.66 | 0.67 | 0.72 |
| HIP_vec_cent | 0.79 | 0.69 | 0.78 | 1.00 | 0.78 | 0.69 | 0.71 | 0.64 | 0.74 | 0.72 | 0.75 |
| HIP_TH_lin | 0.72 | 0.83 | 1.00 | 0.78 | 1.00 | 0.82 | 0.87 | 0.59 | 0.66 | 0.67 | 0.72 |
| HIP_TH_logexp | 0.74 | 0.99 | 0.82 | 0.69 | 0.82 | 1.00 | 0.94 | 0.64 | 0.66 | 0.66 | 0.65 |
| HIP_TH_max | 0.74 | 0.94 | 0.87 | 0.71 | 0.87 | 0.94 | 1.00 | 0.62 | 0.67 | 0.66 | 0.68 |
| HIP_HE_uplift_trad | 0.70 | 0.63 | 0.59 | 0.64 | 0.59 | 0.64 | 0.62 | 1.00 | 0.73 | 0.75 | 0.61 |
| HIP_HE_upproj_trad | 0.75 | 0.66 | 0.66 | 0.74 | 0.66 | 0.66 | 0.67 | 0.73 | 1.00 | 0.70 | 0.79 |
| HIP_HE_uplift_adj | 0.68 | 0.66 | 0.67 | 0.72 | 0.67 | 0.66 | 0.66 | 0.75 | 0.70 | 1.00 | 0.72 |
| HIP_HE_upproj_adj | 0.69 | 0.65 | 0.72 | 0.75 | 0.72 | 0.65 | 0.68 | 0.61 | 0.79 | 0.72 | 1.00 |

Note: see Table *2* for a description of the abbreviations corresponding to each centrality measure.

Analogous conclusions can be drawn from the computation of the top-20 overlap among the financial institutions selected by each metric (Figure *8*). On average, all pairs of indicators coincide in 82% of the entities included among the top 20 most central institutions, while the lowest observed level of overlap is 69%.

Figure 8. **Top-20 overlap among financial institutions identified by each centrality measure**

| | BIP_ degree | BIP_ strength | ONE_ eigenvector | HIP_ vec_cent | HIP_TH_ lin | HIP_TH_ logexp | HIP_TH_ max | HIP_HE_ uplift_trad | HIP_HE_ upproj_trad | HIP_HE_ uplift_adj | HIP_HE_ upproj_adj |
|---|---|---|---|---|---|---|---|---|---|---|---|
| BIP_degree | 1.00 | 0.86 | 0.79 | 0.94 | 0.79 | 0.86 | 0.86 | 0.81 | 0.85 | 0.76 | 0.76 |
| BIP_strength | 0.86 | 1.00 | 0.90 | 0.86 | 0.90 | 0.97 | 0.98 | 0.76 | 0.80 | 0.77 | 0.79 |
| ONE_eigenvector | 0.79 | 0.90 | 1.00 | 0.82 | 1.00 | 0.88 | 0.91 | 0.69 | 0.73 | 0.78 | 0.79 |
| HIP_vec_cent | 0.94 | 0.86 | 0.82 | 1.00 | 0.82 | 0.87 | 0.86 | 0.84 | 0.88 | 0.80 | 0.81 |
| HIP_TH_lin | 0.79 | 0.90 | 1.00 | 0.82 | 1.00 | 0.88 | 0.91 | 0.69 | 0.73 | 0.78 | 0.79 |
| HIP_TH_logexp | 0.86 | 0.97 | 0.88 | 0.87 | 0.88 | 1.00 | 0.96 | 0.77 | 0.80 | 0.77 | 0.79 |
| HIP_TH_max | 0.86 | 0.98 | 0.91 | 0.86 | 0.91 | 0.96 | 1.00 | 0.76 | 0.80 | 0.77 | 0.79 |
| HIP_HE_uplift_trad | 0.81 | 0.76 | 0.69 | 0.84 | 0.69 | 0.77 | 0.76 | 1.00 | 0.96 | 0.69 | 0.71 |
| HIP_HE_upproj_trad | 0.85 | 0.80 | 0.73 | 0.88 | 0.73 | 0.80 | 0.80 | 0.96 | 1.00 | 0.72 | 0.74 |
| HIP_HE_uplift_adj | 0.76 | 0.77 | 0.78 | 0.80 | 0.78 | 0.77 | 0.77 | 0.69 | 0.72 | 1.00 | 0.84 |
| HIP_HE_upproj_adj | 0.76 | 0.79 | 0.79 | 0.81 | 0.79 | 0.79 | 0.79 | 0.71 | 0.74 | 0.84 | 1.00 |

Note: see Table *2* for a description of the abbreviations corresponding to each centrality measure.

## 5. Systemic impact quantification

In this section, we assess the implications of identifying relevant agents in our financial network according to each of the centrality measures described above. To this end, we compute the impact on the network resulting from a distress shock to the top five institutions ranked by each of the eleven metrics considered here (Table *2*). The effect of this shock on the rest of the system is then estimated through a methodology based on the *DebtRank* algorithm developed by Battiston et al. (2012), but adapted to the setting of credit relationships.

Previous studies have already adapted the *DebtRank* algorithm to credit relationships represented as bipartite networks, for instance in the cases of Brazil (Silva et al., 2018) and Japan (Aoyama et al., 2013). We draw on concepts introduced in both applications to adjust the algorithm to credit relationships, while further extending it to the hypergraph setting.

The exercise proposed here consists of the following procedure. For each month, we take the set of institutions ranked in the top-$k$ according to a given centrality measure, with $k = 5$, and simulate an initial shock equivalent to full distress on the credit portfolio of each of these five institutions. Let $H \in \mathbb{R}_+^{N \times E}$ be the weighted incidence matrix of the hypergraph, where $N$ is the number of financial institutions, $E$ is the number of hyperedges, and $H_{ie}$ represents the amount lent by the bank $i$ in hyperedge $e$. The total portfolio of each financial institution is defined as:

$$L_i = \sum_{e=1}^{E} H_{ie}$$

while the total amount associated with each hyperedge is:

$$B_e = \sum_{i=1}^{N} H_{ie}$$

Given the initial set of affected banks $S$, corresponding to the top-$k$ institutions according to the centrality measure under analysis, the initially affected amount of credit in the system is defined as:

$$D_i(0) = \begin{cases} L_i & \text{if } i \in S \\ 0 & \text{if } i \notin S \end{cases}$$

The relative distress level of each institution is then defined:

$$h_i(t) = \frac{D_i(t)}{L_i}$$

bounded in the interval $[0,1]$. By construction, initially shocked institutions have $h_i(0) = 1$. Propagation takes place iteratively through a bank-hyperedge-bank dynamic. Importantly, at each iteration only the new incremental shock generated between $(t-1)$ and $t$ is considered:

$$\Delta h_i(t) = \max\{h_i(t) - h_i(t-1), 0\}$$

This new incremental distress is transmitted from financial institutions to hyperedges according to the relative participation of the affected institutions in each hyperedge:

$$\Delta g_e(t) = \frac{\sum_{i=1}^{N} H_{ie}\, \Delta h_i(t)}{B_e}$$

The term $\Delta g_e(t)$ therefore measures the new fraction of the hyperedge $e$ that becomes affected at iteration $t$. This impact on hyperedges is then transmitted back to financial institutions according to their exposure in each hyperedge:

$$\Delta D_i(t+1) = (1 - h_i(t)) \sum_{e=1}^{E} H_{ie}\, \Delta g_e(t)$$

The factor $(1 - h_i(t))$ prevents an institution from receiving a shock larger than its total portfolio, thereby imposing the upper bound $D_i(t) \leq L_i$. The cumulative affected amount is calculated as follows:

$$D_i(t+1) = \min\{L_i, D_i(t) + \Delta D_i(t+1)\}$$

It is then normalized as a share of each institution's initial total portfolio:

$$h_i(t+1) = \frac{D_i(t+1)}{L_i}.$$

The process continues iteratively until the new aggregate distress in the system converges to a value close to zero (in this application, the algorithms converge on average in 48 iterations per month). Finally, the total systemic impact derived from shocking a set $k$ of banks is defined as:

$$I^{\text{total}} = I^{\text{direct}} + I^{\text{amplified}}$$

Where $I^{\text{direct}} = \sum_{i \in S} L_i$ is the total amount of credit initially affected, and $I^{\text{total}} = \sum_{i=1}^{N} D_i\,(T)$, with $T$ denoting the last iteration. Thus, $I^{\text{total}}$ represents the total systemic impact, in terms of the amount of credit affected, after the iterative propagation dynamics of the initial shock have concluded. The indirect, or amplified, impact is therefore obtained as the difference between these two values. These impacts are reported below as a percentage of the total amount of credit in the network.

In summary, the algorithm measures how much credit in the system becomes affected when the most central banks according to each centrality metric are initially removed or shocked. The hypergraph structure allows propagation to occur not only through bilateral links, but also through co-financing configurations that simultaneously involve multiple financial institutions.

To compare the systemic risk arising from the top five institutions identified by each metric, we focus on the indirect, or amplified, effect rather than on the initial shock, in order to specifically quantify the incremental contagion generated by each set of institutions. This approach is consistent with Aoyama et al (2013), Bardoscia et al (2015) and Silva et al (2018).

Figure 9. **Amplified systemic impact of shocking the top 5 financial institutions identified by each centrality metric** (% of total credit in the system)

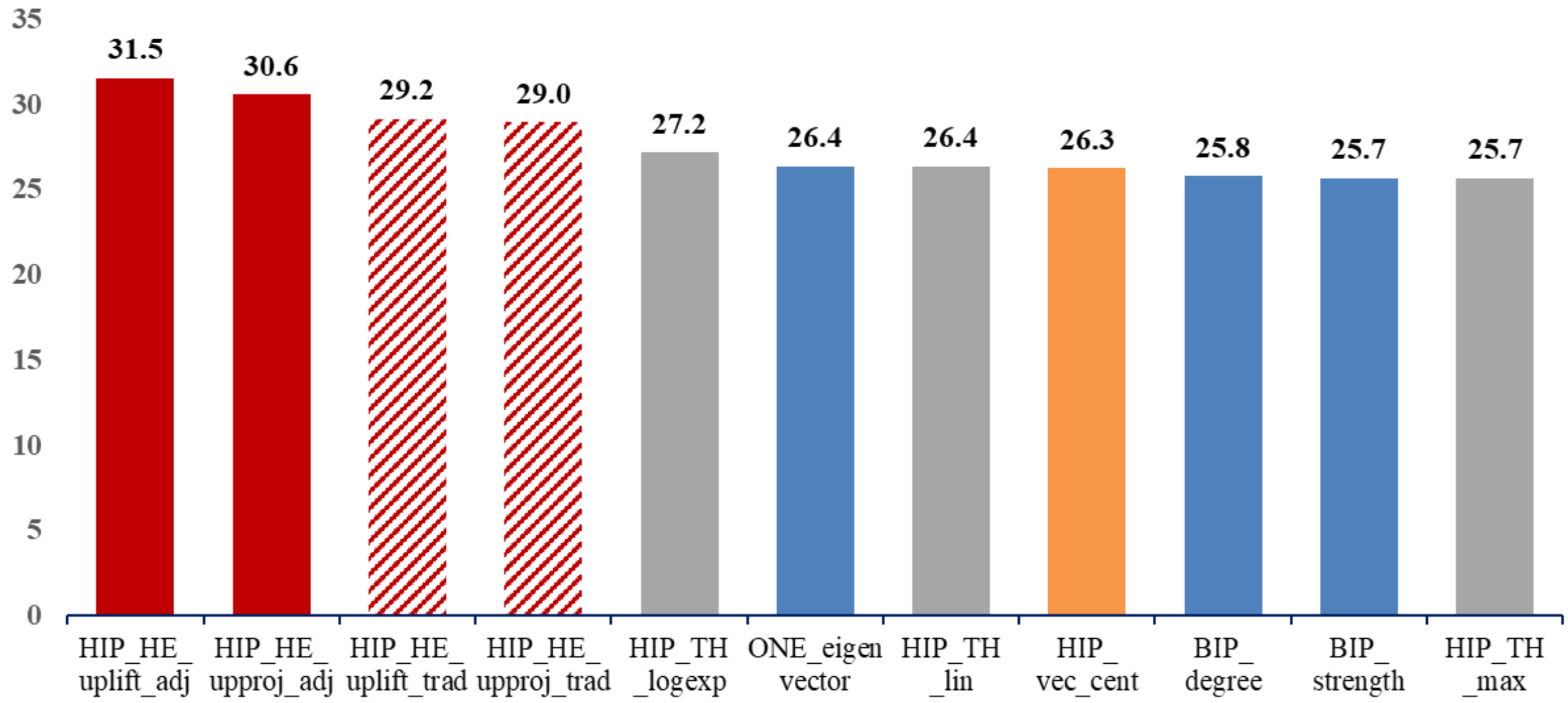

Notes: Solid red columns report the results for the uniformized hypergraph metrics with bank-level individual weighting; red columns with hatching correspond to the uniformized hypergraph metrics with traditional weighting; gray columns correspond to the metrics proposed by Tudisco and Higham (2021); the orange column to vectorial centrality; and blue columns to traditional bipartite-network metrics and their bank-to-bank projection. See Table 2 for a summary of the centrality metrics and their abbreviations. The figure reports the average amplified impact across all months under analysis, although the results remain stable at the monthly level. Appendix C provides further detail on the monthly results.

The comparison of the systemic impact generated by shocks to the top five institutions selected according to each centrality metric (Figure 9, and Appendix C for further detail on the monthly results) yields a clear result: H-eigenvector centrality metrics computed on uniformized hypergraphs identify the set of institutions with the greatest shock-amplification effect. The metrics proposed by Tudisco and Higham and the vectorial centrality measure display more variable results depending on the month and the specific model considered, but they do not consistently outperform traditional metrics such as degree, strength, and/or eigenvector centrality derived from bipartite networks or their one-mode projections.

In particular, the largest amplified impact is generated in every month of the period under analysis by the central agents identified according to the metric proposed in this paper: H-eigenvector centrality computed on hypergraphs uniformized through the methods of Contreras-Aso et al., with centrality scores weighted by the amounts contributed by each bank to each hyperedge (*HIP_HE_ uplift_adj* and *HIP_HE_ upproj_adj*). This metric consistently captures a higher level of systemic risk than all the others. A robustness exercise is presented in Appendix D, where we compute the same systemic impact, but using the original bipartite network (rather than the hypergraph representation), and applying the version of the DebtRank algorithm appropriate to that setting. The results remain virtually identical.

We cannot conclude that there is a single hypergraph uniformization criterion that consistently captures the largest systemic impact. Although, on average, the largest effect is obtained by uniformization through uplifting, in some months the systemic risk detected by the combination of uplifting smaller hyperedges and projecting larger ones to the modal size is higher. However, we can conclude that it is always one of these two metrics (or both) that captures the greatest potential for shock amplification relative to the rest of measures.

This result was, *a priori*, intuitive, since the metric proposed in this paper captures the two most relevant dimensions for quantifying the centrality of a financial institution: on the one hand, how much each institution lends in the network (as a proxy for size), and, simultaneously, the centrality of its neighbors in a setting of multilateral relationships with nonlinear, multiplicative interactions. In other words, it multiplicatively combines an institution's relevance in terms of its lending volume in the network with both its own connectivity and the connectivity of its neighbors, within a framework of multilateral relationships.

## 6. Concluding remarks

This paper proposes the use of hypergraphs to provide a more comprehensive analysis of the structure of credit relationships between financial institutions and firms. In this context, the joint exposures of multiple banks to a firm can be more realistically understood as an underlying multilateral relationship among all institutions lending to that borrower. Representing these relationships as collections of pairwise bilateral interactions may hide relevant information arising from potential amplification effects that simultaneously involve multiple agents.

We compare the results of identifying the most relevant institutions in the network using a range of different centrality measures, from traditional indicators to more recently developed hypergraph-based metrics. We also propose an adjustment to weighted H-eigenvector centrality, after uniformizing hyperedge sizes using the method developed by Contreras-Aso et al. (2024). Our adjustment assigns institution-specific weights based on the amount lent by each institution within each hyperedge.

Based on the central institutions identified by these measures, we quantify the systemic impact of shocking the top five institutions according to each ranking. We find that the greatest amplification effect is generated by distressing the institutions identified by the measure proposed here: H-eigenvector centrality computed on uniformized hypergraphs with banks-specific weights based on the amount lent by each institution within each hyperedge. This method detected entities whose joint shock-amplification effect was, on average, approximately 5 percentage points higher than that associated with the entities selected by traditional metrics (measured as a share of total commercial credit in the system).

These findings have relevant implications for banking supervision and regulation. The proposed measure provides a novel criterion for identifying central institutions and captures greater potential systemic impact than the alternative measures considered. Regulators may therefore find it useful to explore these techniques as a complementary tool to the existing criteria for identifying systemically important banks.

Looking ahead, the research agenda remains fertile, as the application of hypergraphs to financial networks is still at a very early stage. First, this analysis should be extended to other countries and financial networks to assess the external validity of our findings. Other potentially valuable extensions include: a) incorporating the temporal dimension of hypergraphs to study dynamic patterns, b) integrating additional layers of the financial system, or c) developing probabilistic models of agent behavior in an environment characterized by multilateral interactions.

## 7. Appendices

**Appendix A. Node and edge nonlinear eigenvector centrality for hypergraphs (Tudisco & Higham, 2021)**

Let $H = (V, E)$ be a hypergraph with $n$ nodes and $m$ hyperedges. The structure of the hypergraph can be represented by an incidence matrix $B \in \mathbb{R}^{n\times m}$ , whose elements are defined as

$$B_{i,e} = \begin{cases} 1 & \text{if } i \in e, \\ 0 & \text{otherwise} \end{cases}$$

The framework proposed by Tudisco & Higham allows weights to be incorporated for both nodes and hyperedges. To this end, two diagonal matrices are introduced: $N$, associated with node weights, and $W$, associated with hyperedge weights. Node centrality is represented by a positive vector $x \in \mathbb{R}_+^n$, while hyperedge centrality is represented by a positive vector $y \in \mathbb{R}_+^m$. The general formulation consists of solving the following nonlinear system:

$$\lambda x = g\big(BWf(y)\big),$$
$$\mu y = \psi\big(B^T N\varphi(x)\big),$$

where $f$, $g$, $\varphi$ and $\psi$ are functions acting componentwise, and $\lambda, \mu > 0$ are normalization scalars. The first equation indicates that the centrality of each node depends on the centrality of the hyperedges in which it participates. The second equation indicates that the centrality of each hyperedge depends on the centrality of the nodes it contains. Therefore, the measure is defined as a nonlinear eigenvector problem, in which node and hyperedge centralities are jointly determined.

**A1. Linear model.** The first case considered by the authors is obtained by setting

$$f = g = \varphi = \psi = \text{id},$$

where id denotes the identity function. In this case, the previous system reduces to a linear formulation:

$$\lambda x = BWy,$$
$$\mu y = B^T Nx.$$

This specification is analogous to a HITS-type centrality. Nodes define their centrality from the hyperedges to which they belong, while hyperedges centrality is determined by the nodes they contain. In other words, a node is central if it participates in central hyperedges, and a hyperedge is central if it is composed of central nodes. In the case of an ordinary graph, this formulation is related to the computation of eigenvector centrality in the original graph and in the line graph. In hypergraphs, it can be interpreted as a spectral centrality associated with projections of the incidence matrix onto nodes and/or hyperedges.

**A2. Log-Exp Model.** The second model is obtained from the following choice of functions:

$$f(x) = x$$
$$g(x) = \sqrt{x}$$
$$\varphi(x) = \ln(x)$$
$$\psi(x) = \exp(x)$$

This specification introduces a multiplicative form of aggregation. In particular, the centrality of a hyperedge depends on the product of the centralities of the nodes that compose it. For a hyperedge $e$, this yields an expression of the form

$$\mu y_e = \exp\left(\sum_{j\in e} \nu(j)\ln(x_j)\right) = \prod_{j\in e} x_j^{\nu(j)}$$

where $\nu(j)$ represents the weight of node $j$. When node weights are binary, this formulation can be interpreted as a non-uniform generalization of tensor-based Z-eigenvector centralities.

The intuition behind the model is that hyperedges are more central when they are composed of central nodes. However, by relying on multiplicative aggregation, the presence of a node with low centrality significantly reduces the centrality of the entire hyperedge. Therefore, this specification penalizes hyperedges that contain less relevant nodes.

**A3. "Max" Model.** The third model is based on an approximation of the maximum function through a high-order norm. In this case, the authors set

$$f = g = \text{id}$$
$$\varphi(x) = x^{\alpha},$$
$$\psi(x) = x^{1/\alpha}$$

with $\alpha = 10$. The idea behind this specification is that, for high values of $\alpha$,

$$\left(v_1^{\alpha} + \cdots + v_m^{\alpha}\right)^{1/\alpha} \approx \max\{v_1, \ldots, v_m\}$$

In this way, the centrality of a node can be approximated by

$$x_i \approx \max\{y_e : i \in e\}$$

Intuitively, this model assigns high centrality to a node if it participates in at least one highly important hyperedge. Unlike the linear model, it does not add up the contribution of all hyperedges in which the node participates; and unlike the log-exp model, it does not require all those hyperedges to be composed of central nodes. Instead, it approximates a logic of "maximum exposure": the importance of the node is determined primarily by its participation in the most relevant hyperedge.

**A4. Non-linear Power Method.** From a computational perspective, Tudisco & Higham show that, under certain homogeneity and monotonicity conditions on the functions involved, the problem admits a unique positive solution, up to scaling. This solution can be approximated through a nonlinear power method.

The algorithm starts from positive initial vectors for $x$ and $y$, and then proceeds by iteratively alternating between node and hyperedge centrality updates. At each iteration, $x$ is updated on the basis of the current values of $y$, after which $y$ is updated using the newly computed values of $x$. After each step, both vectors are normalized to control for scale. The procedure is repeated until the change between successive iterations becomes sufficiently small.

In applied terms, this approach makes it possible to obtain simultaneously a ranking of nodes and a ranking of hyperedges in large, weighted, non-uniform hypergraphs. This feature is especially useful when the aim is to evaluate the relative importance of nodes not only according to the number of hyperedges in which they participate, but also according to the endogenous importance of those hyperedges and of the other nodes that compose them.

**Appendix B. Uplift and Projection procedures for the uniformization of non-uniform hypergraphs (Contreras-Aso et al., 2024)**

In non-uniform hypergraphs, hyperedges may have different sizes. This entails an obstacle to a direct application of tensor-based spectral measures, such as H-eigenvector centrality, since these measures are naturally formulated for $m$-uniform hypergraphs (that is, hypergraphs in which all hyperedges have the same size). Contreras-Aso et al. propose addressing this problem through a hypergraph uniformization procedure that combines two operations: the uplift of lower-order hyperedges and the projection of higher-order ones. The objective is to transform the original non-uniform hypergraph into a weighted uniform hypergraph, on which a well-defined H-eigenvector centrality can be computed.

Specifically, let $\mathcal{H} = (V, E, w)$ be a weighted hypergraph, where $V$ is the set of nodes, $E$ is the set of hyperedges, and $w(e)$ denotes the weight of hyperedge $e$. Let $\mid e \mid$ reflect the size of hyperedge $e$ and let $M = \max_{e \in E} \mid e \mid$ be the maximum order observed in the hypergraph. To compute a spectral centrality measure, a target order $p$, with $2 \leq p \leq M$, is defined. Then, the procedure transforms all hyperedges so that they have the same size $p$.

**B1. "Uplift" Procedure.** This procedure is applied to hyperedges whose size is smaller than the target order, that is, to every hyperedge $e \in E$ such that $\mid e \mid = k < p$. The key idea is to add an auxiliary node, denoted by $\star$, as many times as necessary for the hyperedge to reach size $p$. Thus, an original hyperedge $e = \{i_1, \dots, i_k\}$ is transformed into:

$$\tilde{e} = \{i_1, \dots, i_k, \underbrace{\star, \dots, \star}_{p-k \text{ times}}\}.$$

The node $\star$ does not belong to the original system, since it is an auxiliary node introduced only to allow for the tensor-based uniformization of the hypergraph. Because the adjacency tensor of an undirected hypergraph is considered symmetric, the introduction of the auxiliary node generates more tensor permutations than those associated with the original hyperedge. For this reason, Contreras-Aso et al. introduce a combinatorial correction factor that avoids overcounting the strength of the original relationship. For a hyperedge of size $k$ increased to order $p$, the corrected tensor weight can be written as

$$\widetilde{w}(e) = w(e) \frac{(p-k)k!}{p!}.$$

This factor preserves the total contribution of the original hyperedge once tensor symmetrization is taken into account. In the simple case of an edge of size 2 lifted to order 3, the factor is

$$\frac{(3-2)2!}{3!} = \frac{1}{3}$$

which is precisely the example discussed by the authors: an interaction $\{i, j\}$, is transformed into $\{i, j, \star\}$, but each artificially added tensor entry receives weight $1/3$ in order to avoid introducing spurious connectivity.

Once the uniform hypergraph has been constructed by increasing the size of smaller hyperedges, *H-eigenvector centrality* is computed on the associated tensor. For a $p$-uniform hypergraph with adjacency tensor $\mathcal{T}$, the H-eigenvector is subject to scaling: if $c$ is a solution, any positive multiple of $c$ is also a solution. Therefore, the centrality of the auxiliary node $\star$ can be formally normalized to $c_\star = 1$. Then, for the empirical analysis, the component associated with the auxiliary node is discarded, and only the centralities of the original nodes in $V$ are maintained. This property is key for making the *uplift* procedure compatible with H-eigenvector centrality in uniform hypergraphs.

**B2. Projection Procedure**. This procedure is applied to hyperedges whose size is larger than the target order, that is, to every hyperedge $e \in E$ with $|e| = k > p$. In this case, the original hyperedge is replaced by all its possible sub-hyperedges of size $p$. Formally, for each $e$, the set $\mathcal{S}_p(e) = \{s \subset e : |s| = p\}$ is considered. Therefore, a hyperedge of size $k$ generates $\binom{k}{p}$ projected hyperedges of size $p$. For example, if $e = \{1,2,3,4\}$ and it is projected to $p = 2$, the six edges $\{1,2\}, \{1,3\}, \{1,4\}, \{2,3\}, \{2,4\}, \{3,4\}$ are obtained. The weighting of the projected hyperedges follows a counting criterion: each sub-hyperedge of size $p$ receives one contribution for each higher-order original hyperedge in which it participates. In the weighted case, this contribution may accumulate the weight $w(e)$ of the original hyperedge. Thus, if several higher-order hyperedges generate the same sub-hyperedge $s$, the final weight of $s$ is obtained by summing those contributions:

$$\widehat{w}(s) = \sum_{e \in E:\, s \subset e,\, |e| > p} w(e).$$

This criterion generalizes the logic of clique projection: a higher-order interaction is represented through all its lower-order interactions, but without introducing factors that artificially increase its relative participation.

**B3. Combination: *Uplifted-Projected* H-Eigenvector Centrality**. The full procedure combines both operations. Given a target order $p$:

$$e \in E \Rightarrow \begin{cases} \textit{uplift} \text{ of } e, & \text{if } |e| < p \\ e \text{ is preserved}, & \text{if } |e| = p \\ \text{projection of } e, & \text{if } |e| > p \end{cases}$$

The result is a weighted $p$-uniform hypergraph. H-eigenvector centrality is then computed on this transformed hypergraph. The resulting measure is referred to by the authors as *$p$-Uplifted-Projected H-Eigenvector Centrality* ($p$-UPHEC). When $p = M$, the procedure is equivalent to lifting all smaller hyperedges to the maximum observed order; in this case, the resulting measure corresponds to $m$-Uplifted H-Eigenvector Centrality ($m$-UHEC). When $p$ is lower than $M$, the procedure combines the projection of larger hyperedges with the *uplift* of smaller ones.

If the original hypergraph is strongly connected, the transformed hypergraph preserves the connectivity required to apply Perron-Frobenius-like associated theorems for nonnegative tensors. Therefore, a positive H-eigenvector centrality exists and is unique (up to scaling). This property allows the resulting vector to be interpreted as a well-defined centrality measure for the original nodes of the hypergraph.

Intuitively, the procedure makes it possible to incorporate information from interactions of different orders simultaneously. Smaller hyperedges are not discarded, but rather lifted through the auxiliary node; larger hyperedges are not eliminated either, but are projected onto the selected order of analysis. In this way, the resulting centrality summarizes the importance of each node within the full structure of the non-uniform hypergraph, avoiding the need to compute separate centralities for each hyperedge order. This is particularly useful in empirical applications where higher-order interactions are heterogeneous, as in financial networks.

**Appendix C. Amplified systemic impact of shocking the top 5 financial institutions identified by each centrality metric** (% of total credit in the system)

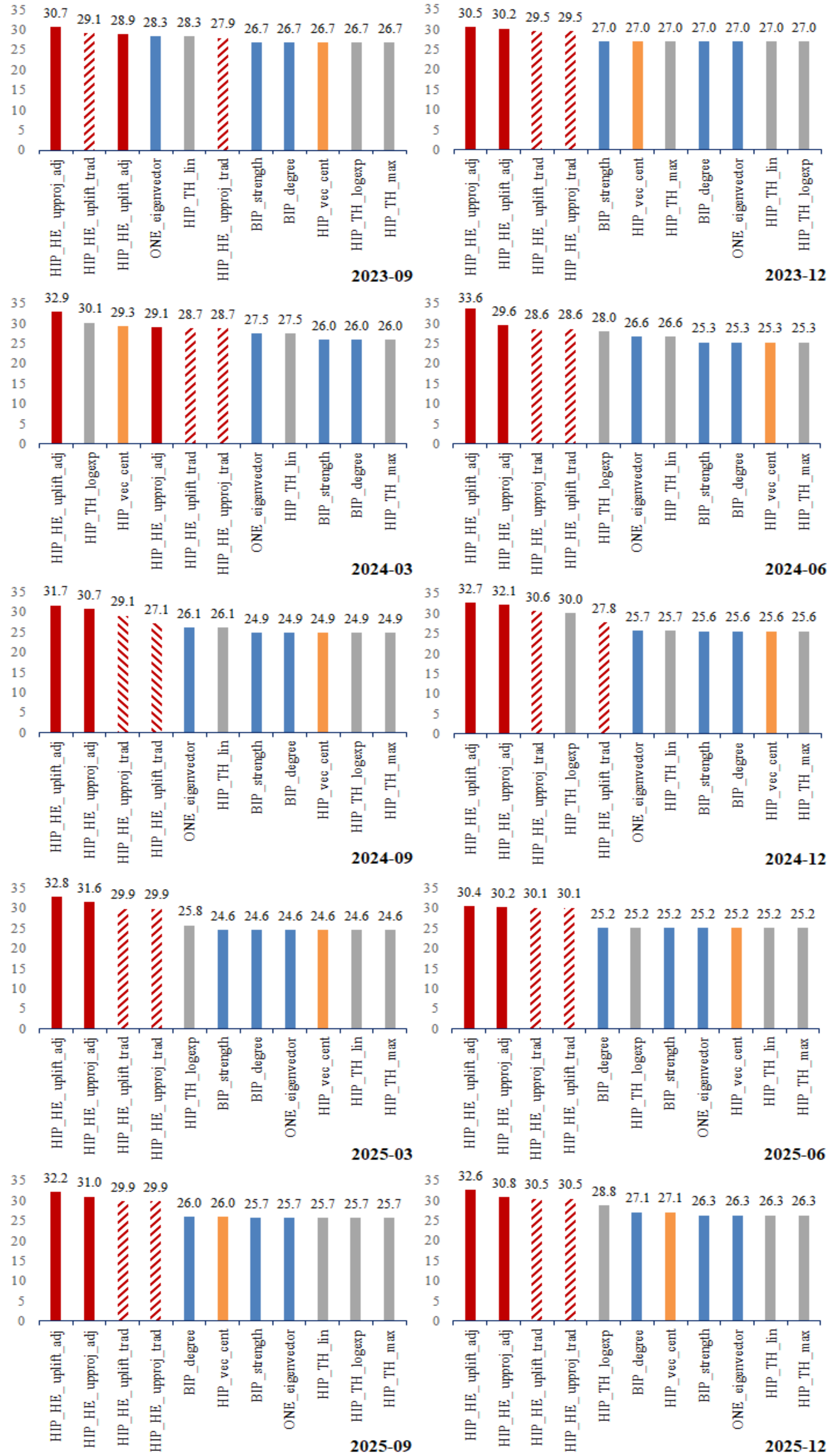

Note: See Figure 9 and Table 2 for further details on the estimated effects and the centrality measures considered. For reasons of space, only quarter-end observations are reported; however, the results are similar for all months in the period under study.

### Appendix D. Robustness check: amplified systemic impact measured on traditional bipartite networks

As a robustness exercise, we computed the systemic impact resulting from the distress of the top-ranked financial institutions using the original bipartite network, rather than the hypergraph representation. To this purpose, we applied a DebtRank algorithm adapted to the bipartite setting, analogous to the one presented above for hypergraphs and consistent with the approaches used by Silva et al. (2018) for the Brazilian credit network and by Aoyama et al. (2013) for the Japanese credit network.

The results (reported in Figure 10) show that both the systemic impact and the relative magnitudes associated with each centrality measure remain virtually identical to those obtained in the main exercise of this paper (reported in Figure 9). This confirms that our conclusions do not depend on the specific DebtRank algorithm implemented for hypergraphs.

Figure 10**. Amplified systemic impact of shocking the top 5 financial institutions identified by each centrality metric on the original bipartite network** (% of total credit in the system)

Notes: Solid red columns report the results for the uniformized hypergraph metrics with bank-level individual weighting; red columns with hatching correspond to the uniformized hypergraph metrics with traditional weighting; gray columns correspond to the metrics proposed by Tudisco and Higham (2021); the orange column to vectorial centrality; and blue columns to traditional bipartite-network metrics and their bank-to-bank projection. See Table 2 for a summary of the centrality metrics and their abbreviations. The figure reports the average amplified impact across all months under analysis, although the results remain stable at the monthly level.